\documentclass[doublecol]{epl2} 

\usepackage[english]{babel}
\usepackage{amsmath}
\usepackage{amssymb}
\usepackage{hyperref}

\makeatletter

\title{Specular transmission and diffuse reflection in phonon scattering
at grain boundary}

\author{Zhun-Yong Ong\inst{1}}

\institute{                    
  \inst{1} Institute of High Performance Computing, A{*}STAR, Singapore 138632,
Singapore
}
\pacs{63.20.kp}{Phonon-defect interactions}

\abstract{
It is widely assumed in the literature that the specularity parameters
for phonon transmission (forward scattering) and reflection (backward
scattering) at a boundary are identical, i.e., the statistical distributions
of the transition probabilities between an incident phonon and the
range of outgoing phonon modes are the same for both transmission
and reflection. However, it is hypothesized by Li and McGaughey that
separate specularity parameters are needed to describe the behavior
of transmitted and reflected phonons in superlattices and polycrystalline
materials correctly. We test this hypothesis by analyzing the mode-resolved
specularity parameters computed separately for transmission and reflection
processes at a graphene grain boundary. Our results show that backward
scattering is considerably more diffuse than forward scattering at
most frequencies and polarizations, providing strong evidence for
Li and McGaughey's hypothesis, and shed new light on how surfaces
and interfaces modify phonon transport within and between domains
in nanostructured materials.
}

\begin{document}

\maketitle

\section{Introduction}
One of the main factors limiting the
lattice thermal conductivity of nanostructures (e.g. Si nanowires~\cite{AHochbaum:Nature08_Thermoelectric,JLim:NL12_Quantifying})
and polycrystalline materials is boundary scattering at free surfaces
and solid-solid interfaces~\cite{DLi:NMTE15_Phonon,CMonachon:ARMR16_Thermal},
a process which impedes phonon propagation by dissipating phonon momentum.
The degree of momentum dissipation per scattering event depends on
its specularity and is characterized by the specularity parameter
($\mathcal{P}$) which satisfies the condition $0\leq\mathcal{P}\leq1$
and varies with boundary morphology as well as the momentum, frequency
and branch of the incident phonon~\cite{AMaznev:PRB15_Boundary,NRavichandran:PRX18_Spectrally}. 

In perfectly specular scattering ($\mathcal{P}=1$), the incident
phonon is either reflected or transmitted specularly with no loss
of transverse momentum (momentum in the direction parallel to the
boundary) while in perfectly diffusive scattering ($\mathcal{P}=0$)
or the so-called \emph{Casimir} limit, the incoming phonon energy
is redistributed uniformly over the entire spectrum of outgoing phonon
channels with total loss of transverse momentum 
due to randomization
of the phonon trajectory~\cite{GChen:Book05_Nanoscale}. In reality,
most boundary scattering processes are neither fully specular nor
diffuse, i.e., $0<\mathcal{P}<1$, even for a rough boundary because
the scattering amplitudes, which determine the transitions between
the incoming and outgoing phonon states, have a nonuniform distribution.
Hence, the determination of the specularity parameters can provide
an accurate quantification of the momentum loss and thermal conductivity
reduction from boundary scattering.

In the case of a solid-solid interface, a phonon is partially forward-scattered
(i.e. transmitted) across the boundary and partially backward-scattered
(i.e. reflected) into the original bulk lattice, resulting in thermal
resistance in the directions normal and parallel to the boundary.
In the normal direction, momentum is lost directly as a result of
incomplete transmission across the interface while in the parallel
direction, transverse momentum is lost through diffuse scattering
at the boundary. In superlattices where there are multiple interfaces~\cite{GChen:PRB98_Thermal,KTermentzidis:PRB09_Nonequilibrium},
the first process creates thermal resistance in the cross-plane direction
while the second process reduces thermal conductivity in the in-plane
direction.

\begin{figure}
\begin{centering}
\includegraphics[scale=0.35]{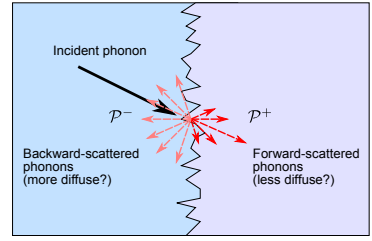}
\par\end{centering}
\caption{Illustration of the transmission and reflection specularity parameters
($\mathcal{P}^{+}$ and $\mathcal{P}^{-}$) for phonon scattering
at a solid-solid interface. $\mathcal{P}^{+}$ and $\mathcal{P}^{-}$
measure the degree of randomization in the forward and backward-scattered
phonons, respectively.}

\label{fig:ScatteringSpecularityCartoon}
\end{figure}

It is widely assumed that in the second process, transverse momentum
dissipation is similar for phonon transmission and reflection~\cite{GChen:Book05_Nanoscale,HZhao:JAP09_Phonon,ZAksamija:PRB13_Thermal,QHao:JAP14_General,CHua:SST14_Grain,YZhang:FER18_Modified}
so that the degree of randomization for transmitted phonons is the
same as that for reflected phonons. In many models where the scattering
is only partially random,~\cite{HZhao:JAP09_Phonon,ZAksamija:PRB13_Thermal,QHao:JAP14_General,CHua:SST14_Grain,YZhang:FER18_Modified}
the same specularity parameter is used to determine the phonon distributions
on both sides of the interface. In the diffuse mismatch model,~\cite{ETSwartz:RMP89_Thermal}
the transmitted and reflected phonons are completely randomized.
However,
it has been suggested by Li and McGaughey~\cite{DLi:NMTE15_Phonon}
that phonon reflection is much more diffuse and sensitive to boundary
roughness than phonon transmission, i.e., the incident phonon is spread
more diffusely over the outgoing (reflected) phonon channels on the
same side of the boundary than it is over the outgoing (transmitted)
phonon channels on the other side as depicted in Fig.~\ref{fig:ScatteringSpecularityCartoon},
and thus, greater transverse momentum dissipation results from reflection
than transmission. This difference in the diffuseness of scattering
draws further support from Ref.~\cite{FShi:PRB17_Diffusely} where
Shi, Lowe and Craster analyzed the elastic scattering of waves in
an elastodynamic model of the Si-Ge interface using the Kirchhoff
approximation~\cite{EThorsos:JASA88_Validity} (KA) and concluded
within the validity of the KA and the continuum limit that the reflection
and transmission of low-frequency (sub-THz) waves can have different
dependence on interface roughness when the correlation length of the
boundary is large. Therefore, the directional dependence of boundary
scattering implies that two specularity parameters, one for reflection
($\mathcal{P}^{-}$) and the other for transmission ($\mathcal{P}^{+}$),
are needed to describe phonon scattering at the interface of two media.

If proven to be true, this distinction between the reflection and
transmission specularity can potentially resolve some of the puzzles
in nanoscale thermal transport involving phonon interactions with
surfaces and interfaces~\cite{DLi:NMTE15_Phonon}. For example,
a specularity of $\mathcal{P}\approx0$ is needed to interpret the
experimental characterization of thermal transport in freestanding
nanowires and thin films which have only phonon reflection from free
surfaces~\cite{DLi:APL03_SiNW,DLi:APL03_SiGeNW} while a significantly
larger average specularity parameter is needed to model the in-plane
thermal conductivity of superlattices where there is both phonon transmission
and reflection~\cite{GChen:JHT97_Size}, a discrepancy that can
be explained by assuming distinct specularity parameters for reflection
and transmission and that the transmission specularity is greater
than the reflection specularity, i.e. the transmitted phonons are
less randomized than the reflected phonons ($\mathcal{P}^{+}\gg\mathcal{P}^{-}$).
However, the evidence supporting distinct specularity parameters for
phonon reflection and transmission remains only circumstantial as
it is based on experimental data interpretation~\cite{DLi:NMTE15_Phonon}.
Even in simulations, the direct evaluation of $\mathcal{P}^{+}$ and
$\mathcal{P}^{-}$, which are not defined precisely for individual
phonon modes, is constrained by the difficulty in computing the boundary
scattering amplitudes. 

In this article, we show that phonon transmission is less diffuse
than phonon reflection over most of the Brillouin zone (BZ), highlighting
the need for distinct specularity parameters to characterize phonon
reflection and transmission. To accomplish this, we define and directly
evaluate the \emph{mode-resolved} specularity parameters for phonon
reflection ($\mathcal{P}^{-}$) and transmission ($\mathcal{P}^{+}$)
at a (32,32)$|$(56,0) GB between armchair and zigzag-edge graphene~\cite{GSchusteritsch:PRB14_Predicting}.
Our approach, which is facilitated by the recently developed $S$-matrix
method~\cite{ZYOng:PRB18_Atomistic} for calculating the \emph{exact}
scattering amplitudes, is based on the analysis of the statistical
spread of transition probabilities and is elaborated in an earlier
paper~\cite{ZYOng:PRB20_Structure} where the total specularity
($\mathcal{P}$) of the (32,32)$|$(56,0) GB is studied. We choose this
GB structure as our model system because armchair and zigzag-edge
graphene have a similar but not necessarily identical number of incoming
phonon channels at most frequencies, i.e., $N_{\text{AC}}(\omega)\approx N_{\text{ZZ}}(\omega)$
where $N_{\text{AC}}$ ($N_{\text{ZZ}}$) is the number of incoming
phonon channels in armchair-edge (zigzag-edge) graphene at frequency
$\omega$, allowing us to exclude this as a factor affecting the specularity.
In our analysis of the scattering amplitudes, we introduce an expression
for estimating the normalized transmission and reflection specularity
parameters ($P_{n}^{+}$ and $P_{n}^{-}$) of the $n$-th incoming
phonon ($n=1,\ldots,N$ for $N=N_{\text{AC}}+N_{\text{ZZ}}$) and
discuss how $P_{n}^{+}$ and $P_{n}^{-}$ vary with phonon frequency
($\omega_{n}$), momentum ($\boldsymbol{k}_{n}$) and branch ($\nu_{n}$).
We also compute the mode-resolved transmission probability ($\mathcal{T}_{n}$)
and show that its distribution in the BZ is highly similar to that
of the transmission specularity $P_{n}^{+}$.

\section{Graphene grain boundary and $S$-matrix calculation}
\begin{figure}
\begin{centering}
\includegraphics[scale=0.35]{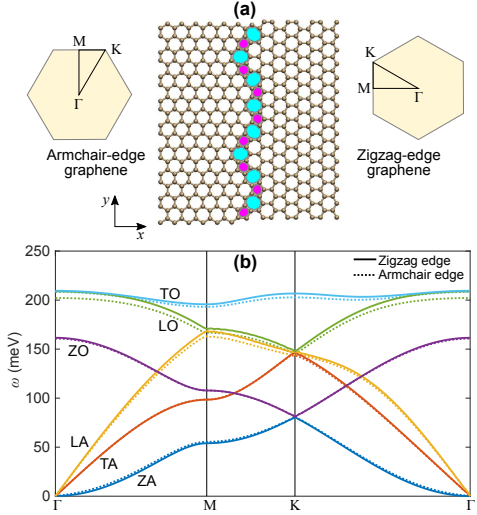}
\par\end{centering}
\caption{\textbf{(a)} Depiction of the GB between armchair and zigzag-edge
graphene. The different graphene lattices (armchair-edge vs. zigzag-edge)
are referred according to their edge orientation in the $y$ direction.
\textbf{(b)} Bulk phonon dispersion curves of armchair (dotted lines)
and zigzag-edge (solid lines) graphene along the major BZ symmetry
points and labeled according to branch~\cite{CGan:JPC21_Complementary}.
The small differences in the phonon dispersion are due to the tensile
strain in armchair-edge graphene in the $y$ direction. Both the TA
and ZA phonon dispersion curves have an inverted Dirac cone at the
$K$ point.}
\label{fig:GBStructurePhononDispersion}

\end{figure}

Although
the details of the generation and optimization of the graphene GB
structures in our simulation are identical to those in Ref.~\cite{ZYOng:PRB20_Structure},
we give a brief overview of the method here with further details in
the Supplemental Material~\cite{Supplemental_Material}. We construct
the (32,32)$|$(56,0) graphene GB model from the two lowest-energy (4,4)$|$(7,0)
GB configurations (GB-II and GB-III) in Ref.~\cite{GSchusteritsch:PRB14_Predicting}
found using the \emph{ab initio} random structure searching method~\cite{CPickard:PRL06_High}.
Each (32,32)$|$(56,0) GB configuration comprises $l=8$ (4,4)$|$(7,0)
GB's, a permutation of GB-II's and GB-III's, forming a continuous
line of pentagon-heptagon defect pairs as shown in Fig.~\ref{fig:GBStructurePhononDispersion}(a).
In total, there are $2^{l}=256$ unique GB configurations. We set
the direction of the phonon flux and the GB to be parallel to the
$x$ and $y$ axis, respectively, and impose periodic boundary conditions
in the $y$ direction, resulting in the discretization of the transverse
component of the wave vector ($k_{y}$) with the step size of $\Delta k_{y}=\frac{2\pi}{L_{y}}$
where $L_{y}$ is the width of the GB. We note here that the number
of modes is finite because of the nonzero $\Delta k_{y}$ associated
with the finite GB width. The empirical Tersoff potential~\cite{JTersoff:PRL88_Empirical},
with parameters from Ref.~\cite{LLindsay:PRB10_Optimized}, is used
to model the C-C interatomic forces. The program GULP~\cite{JGale:MolSim03_gulp}
is used to optimize each GB configuration and to generate its interatomic
force constant (IFC) matrices $\boldsymbol{H}_{\text{CL}}$, $\boldsymbol{H}_{\text{C}}$
and $\boldsymbol{H}_{\text{CR}}$ needed for the $S$-matrix calculations.
We also compute the IFC matrices $\boldsymbol{H}_{\text{L}}^{00}$
and $\boldsymbol{H}_{\text{L}}^{01}$ ($\boldsymbol{H}_{\text{R}}^{00}$
and $\boldsymbol{H}_{\text{R}}^{01}$) describing the bulk phonons
in the armchair-edge (zigzag-edge) graphene on the left (right) side
of the GB, with the phonon dispersion curves along the high-symmetry
directions shown in Fig.~\ref{fig:GBStructurePhononDispersion}(b).

At each frequency $\omega$, we compute an $N(\omega)\times N(\omega)$
matrix $\boldsymbol{S}(\omega)$ for each GB configuration, where
$N(\omega)=N_{\text{AC}}(\omega)+N_{\text{ZZ}}(\omega)$ is the number
of phonon scattering channels, using the IFC matrices $\boldsymbol{H}_{\text{CL}}$,
$\boldsymbol{H}_{\text{C}}$, $\boldsymbol{H}_{\text{CR}}$, $\boldsymbol{H}_{\text{L}}^{00}$,
$\boldsymbol{H}_{\text{L}}^{01}$, $\boldsymbol{H}_{\text{R}}^{00}$
and $\boldsymbol{H}_{\text{R}}^{01}$, based on the method described
in Refs.~\cite{ZYOng:PRB18_Atomistic,ZYOng:PRB20_Structure}. Details
of the calculation of the $S$ matrix are given in the Supplemental
Material~\cite{Supplemental_Material}. The individual $S$-matrix
element $[\boldsymbol{S}(\omega)]_{mn}=S(\boldsymbol{k}_{m},\boldsymbol{k}_{n}^{\prime})$
gives the scattering amplitude from the incoming phonon channel with
wave vector $\boldsymbol{k}_{n}^{\prime}$ and branch index $\nu_{n}^{\prime}$
to the outgoing phonon channel with wave vector $\boldsymbol{k}_{m}$
and branch index $\nu_{m}$. The transition probability matrix $\boldsymbol{W}(\omega)$
is obtained by taking the configurational ensemble average, taken
over all 256 possible GB configurations and denoted by $\langle\ldots\rangle$,
of the absolute square of the $S$-matrix elements~\cite{ZYOng:PRB20_Structure},
i.e., $[\boldsymbol{W}(\omega)]_{mn}=\langle|[\boldsymbol{S}(\omega)]_{mn}|^{2}\rangle$.

\section{Definition of specularity parameters and transmission coefficient}
From
$\boldsymbol{W}(\omega)$, we estimate the normalized total specularity
parameter of the the $n$-th incoming phonon~\cite{ZYOng:PRB20_Structure}
using the expression:
\begin{equation}
P_{n}^{\text{total}}(\omega)=\frac{\sqrt{\sum_{k_{y}}\phi_{n}(k_{y})^{2}}}{\sum_{k_{y}}\phi_{n}(k_{y})}\ ,\label{eq:TotalSpecularity}
\end{equation}
where $k_{y}$ is the discretized transverse wave vector and $\phi_{n}(k_{y})=\sum_{m=1}^{N(\omega)}[\boldsymbol{W}(\omega)]_{mn}\delta_{k_{y,m},k_{y}}$
is the sum of the probability of the $n$-th incoming phonon mode
scattering to outgoing channels with the transverse wave vector $k_{y}$.
Equation (\ref{eq:TotalSpecularity}) is analogous to the inverse
participation ratio used in localization theory~\cite{CMonthus:PRB10_ParticipRatios}
but measures the localization of scattering in $k_{y}$ instead of
real space. In perfectly specular scattering, $k_{y}$ is conserved
so that $\phi_{n}(k_{y})=1$ for $k_{y}=k_{k,n}^{\prime}$ and $0$
otherwise. Hence, Eq. (\ref{eq:TotalSpecularity}) yields $P_{n}^{\text{total}}(\omega)=1$
for perfectly specular scattering and $\lim_{N(\omega)\rightarrow\infty}P_{n}^{\text{total}}(\omega)=0$
for totally diffuse scattering. By grouping together the probabilities
of the outgoing channels with the transverse wave vector $k_{y}$
in $\phi_{n}(k_{y})$, we account for the effects of mode conversion
in specular scattering.

The transmission coefficient or probability~\cite{ZYOng:JAP18_Tutorial}
of the $n$-th incident phonon mode, equal to the sum of its transition
probabilities to outgoing forward-scattered phonon modes, is given
by
\begin{equation}
\mathcal{T}_{n}(\omega)=\sum_{m=1}^{N(\omega)}[\boldsymbol{W}(\omega)]_{mn}\Theta(v_{x,m}v_{x,n}^{\prime})\label{eq:TransmissionProbability}
\end{equation}
where $v_{x,n}^{\prime}$ ($v_{x,m}$) denotes the group velocity
component in the $x$-direction of the $n$-th incoming ($m$-th outgoing)
phonon mode and $\Theta(\ldots)$ represents the Heaviside step function.
The $\Theta(v_{x,m}v_{x,n}^{\prime})$ term in the summand in Eq.~(\ref{eq:TransmissionProbability}),
which equals unity (zero) for outgoing phonon channels associated
with forward (backward) scattering, eliminates the transition probability
contributions from phonon reflection in Eq.~(\ref{eq:TransmissionProbability}).
We use the Heaviside step function in a similar manner in our definition
of the normalized transmission and reflection specularity parameters
($P_{n}^{+}$ and $P_{n}^{-}$): 
\begin{equation}
P_{n}^{\text{\ensuremath{\pm}}}(\omega)=\begin{cases}
\begin{array}{l}
\frac{\sqrt{\sum_{k_{y}}\phi_{n}^{\pm}(k_{y})^{2}}}{\sum_{k_{y}}\phi_{n}^{\pm}(k_{y})}\\
0
\end{array} & \begin{array}{l}
\text{if }\sum_{k_{y}}\phi_{n}^{\pm}(k_{y})>0\\
\text{otherwise}
\end{array}\end{cases}\label{eq:ForwardBackwardSpecularity}
\end{equation}
where $\phi_{n}^{\pm}(k_{y})=\sum_{m=1}^{N(\omega)}[\boldsymbol{W}(\omega)]_{mn}\delta_{k_{y,m},k_{y}}\Theta(\pm v_{x,m}v_{x,n}^{\prime})$
is the sum of the probability of the $n$-th incoming phonon mode
scattering to outgoing channels with the transverse wave vector $k_{y}$
in the forward ($+$) or backward ($-$) direction, so that $\phi_{n}(k_{y})=\phi_{n}^{+}(k_{y})+\phi_{n}^{-}(k_{y})$.
The denominator in Eq.~(\ref{eq:ForwardBackwardSpecularity}) corresponds
to the total transmission ($+)$ or reflection ($-$) probability.
Equation~(\ref{eq:ForwardBackwardSpecularity}) measures the scattering
localization of the transmitted or reflected phonon modes, with $P_{n}^{\text{\ensuremath{+}}}(\omega)=1$
($P_{n}^{\text{-}}(\omega)=1$) for specular transmission (reflection),
i.e., if the forward (backward) scattering processes conserve the
transverse wave vector, and $\lim_{N(\omega)\rightarrow\infty}P_{n}^{\text{\ensuremath{+}}}(\omega)=0$
($\lim_{N(\omega)\rightarrow\infty}P_{n}^{\text{-}}(\omega)=0$) if
the forward (backward) scattering process is totally diffuse. Unlike
Eq.~(\ref{eq:TotalSpecularity}), $P_{n}^{\text{\ensuremath{\pm}}}(\omega)$
allows us to distinguish between the degree of specularity in transmission
(forward scattering) and reflection (backward scattering).

\section{Relationship between transmision probability and total specularity
parameter}
\begin{figure*}
\begin{centering}
\includegraphics[scale=0.23]{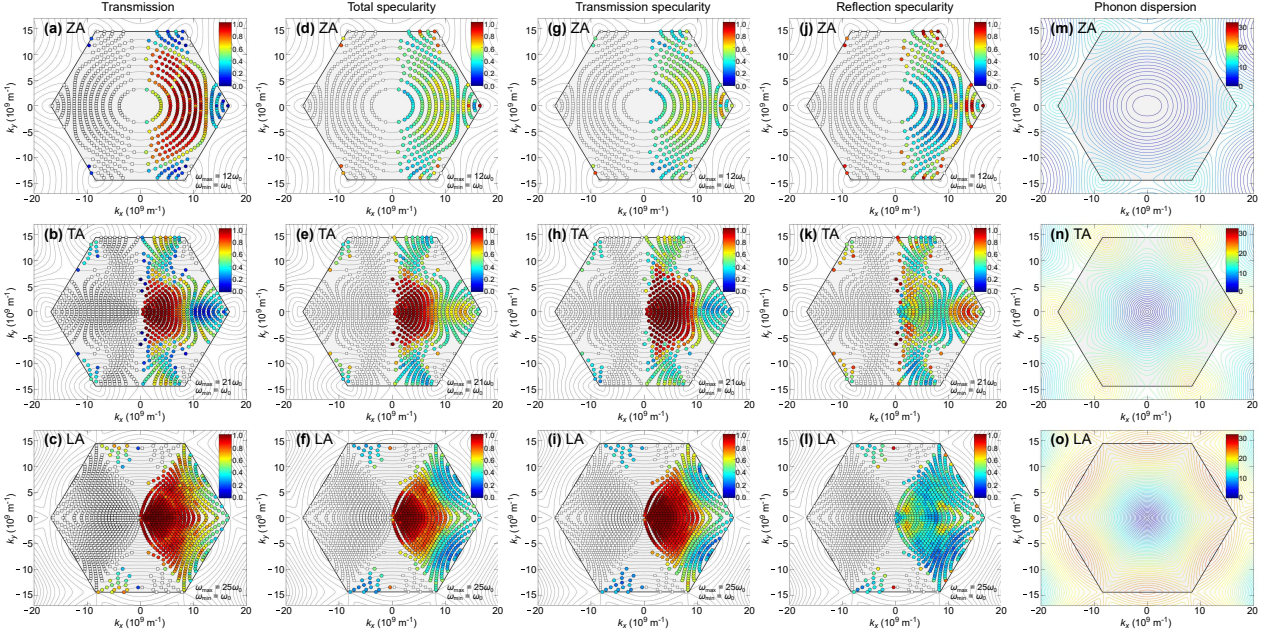}
\par\end{centering}
\caption{Mode-resolved comparison of \textbf{(a-c)} $\mathcal{T}_{n}(\omega)$,
\textbf{(d-f)} $P_{n}^{\text{total}}(\omega)$, \textbf{(g-i)} $P_{n}^{+}(\omega)$,
and \textbf{(j-l)} $P_{n}^{-}(\omega)$ for ZA, TA and LA phonons
in armchair-edge graphene at each frequency $\omega$ point over the
frequency range $\omega=m\omega_{0}$ for $m=$ 1 to 25 and $\omega_{0}=10^{13}$
rad/s ($6.58$ meV). The isofrequency contours at each $\omega$ are
shown using solid gray lines. The modes in the incoming phonon flux
are represented by filled circles, colored according to their numerical
value as indicated in the color bars, while the modes in the outgoing
flux are represented by hollow squares. The \textbf{(m)} ZA, \textbf{(n)}
TA and \textbf{(o)} LA phonon dispersions are indicated with color
contours in intervals of $\Delta\omega=\omega_{0}/2$.}

\label{fig:ModeResolvedTransmissionTotalSpecularity}
\end{figure*}

We compute $\mathcal{T}_{n}$, $P_{n}^{\text{total}}$,
$P_{n}^{+}$ and $P_{n}^{-}$ from Eqs.~(\ref{eq:TotalSpecularity})-(\ref{eq:ForwardBackwardSpecularity})
over the discrete frequency range $\omega=m\omega_{0}$ for $m=1$
to $25$ and $\omega_{0}=10^{13}$ rad/s ($6.58$ meV) in armchair-edge
graphene. At each frequency point $\omega$, we have a set of $N(\omega)$
bulk phonon modes corresponding to the incoming phonon channels, which
we enumerate from $n=1$ to $N(\omega)$. Thus, by sweeping over the
entire frequency range, we obtain the mode-resolved transmission and
specularity parameters over the entire Brillouin zone (BZ) in Fig.~\ref{fig:ModeResolvedTransmissionTotalSpecularity}.
We associate with the $n$-th phonon mode a 2-dimensional wave vector
$\boldsymbol{k}_{n}$ and a branch index $\nu_{n}$ which can be flexural
acoustic (ZA), transverse acoustic (TA) or longitudinal acoustic (LA).
The optical phonon branches are ignored in our study while the corresponding
results for zigzag-edge graphene are given in the Supplemental Material~\cite{Supplemental_Material}.

Figures~\ref{fig:ModeResolvedTransmissionTotalSpecularity}(a) to
(c) show the mode-resolved $\mathcal{T}_{n}$ distribution in the
first BZ for each phonon branch (ZA, TA or LA). We observe that $\mathcal{T}_{n}$
is near-isotropic, showing no significant directional dependence for
all branches except for modes that are close to the BZ edge. In spite
of the 30-degree tilt between armchair- and zigzag-edge graphene,
there is no critical angle for transmission and total internal reflection
like that exhibited for interfaces between two dissimilar materials
(e.g. the graphene/\emph{h}-BN interface~\cite{ZYOng:PRB15_Efficient}).
The near-isotropic $\mathcal{T}_{n}$ distribution and absence of
total internal reflection are attributed to the similar acoustic impedance
between armchair- and zigzag-edge graphene in the continuum (long-wavelength)
limit and are consistent with the acoustic mismatch model (AMM)~\cite{ETSwartz:RMP89_Thermal}
which predicts that $\mathcal{T}_{n}\sim1$ in the continuum (long-wavelength)
limit for the acoustically similar materials.

\begin{figure}
\begin{centering}
\includegraphics[scale=0.28]{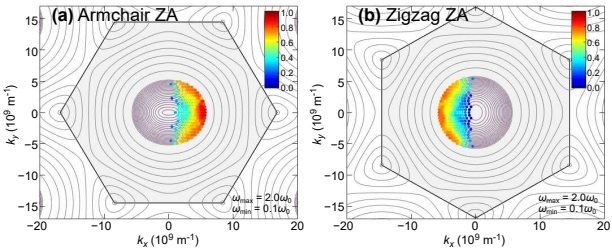}
\par\end{centering}
\caption{Mode-resolved comparison of $\mathcal{T}_{n}(\omega)$ in the low-frequency
regime for ZA phonons in \textbf{(a)} armchair and \textbf{(b)} zigzag-edge
graphene at each frequency $\omega$ point over the frequency range
$\omega=0.1\omega_{0}$ to $2\omega_{0}$ for $\omega_{0}=10^{13}$
rad/s ($6.58$ meV) with a frequency step size of $0.1\omega_{0}$.
The isofrequency contours at each $\omega$ at and under $\omega=2\omega_{0}$
are displayed using solid magenta lines at intervals of $\Delta\omega=0.1\omega_{0}$
while the isofrequency contours between $\omega=2\omega_{0}$ and
$\omega=12\omega_{0}$ are displayed using solid gray lines at intervals
of $\Delta\omega=\omega_{0}$. The modes in the incoming phonon flux
are represented by filled circles, colored according to their numerical
value as indicated in the color bars.}
\label{fig:ModeResolvedLowFreqZATransmission}
\end{figure}

However, we also observe discrepancies with the AMM. In Figs.~\ref{fig:ModeResolvedTransmissionTotalSpecularity}(a)
to (c), the ZA phonon $\mathcal{T}_{n}$ decreases as we approach
the BZ center or edges while the TA and LA phonon $\mathcal{T}_{n}$
converges to unity at the BZ center and decreases as one approaches
the edges. The increasing LA and TA phonon transparency in the $\omega\rightarrow0$
limit is consistent with the AMM and agrees with the findings from
wave-packet simulations~\cite{PKSchelling:APL02_Phonon,PKSchelling:JAP04_Kapitza}
that long-wavelength phonons are more easily transmitted as they are
less sensitive to the crystallographic discontinuity. On the other
hand, the reduced $\mathcal{T}_{n}$ for long-wavelength ZA phonons
suggests that they may be more easily scattered by the GB. 
To investigate
this further, we plot $\mathcal{T}_{n}$ of the low-frequency ZA phonons
over the frequency range of $\omega=0.1\omega_{0}$ to $2\omega_{0}$
for armchair and zigzag-edge graphene in Fig.~\ref{fig:ModeResolvedLowFreqZATransmission}.
The results agree with our findings at high frequencies showing $\mathcal{T}_{n}$
decreasing with frequency and may be connected to the more pronounced
anisotropy of the ZA phonon dispersion in armchair-edge graphene,
which is strained in the $y$ direction, relative to that in zigzag-edge
graphene. As a result of this low-frequency anisotropy, there is an
imbalance in the number of ZA phonon modes between the armchair and
zigzag-edge graphene at low frequencies, with proportionally fewer
modes in the former than in the latter. For instance, at $\omega=0.1\omega_{0}$
the lowest simulated frequency, there is only one available ZA phonon
mode in armchair-edge graphene and five in zigzag-edge graphene. We
have also performed the same calculations for a more structurally
ordered grain boundary constructed from GB-II subunits and the results,
shown in the Supplementary Material,~\cite{Supplemental_Material}
also exhibit a qualitatively similar decrease in $\mathcal{T}_{n}$
with frequency.

To quantify scattering by the GB, we plot in Figs.~\ref{fig:ModeResolvedTransmissionTotalSpecularity}(d)
to (f) the mode-resolved total specularity $P_{n}^{\text{total}}$
distribution {[}Eq.~(\ref{eq:TotalSpecularity}){]} in the first
Brillouin Zone (BZ) for each phonon polarization (ZA, TA or LA). We
notice immediately a close similarity between the $\mathcal{T}_{n}$
and $P_{n}^{\text{total}}$ distribution for the TA and LA phonons:
$\mathcal{T}_{n}$ and $P_{n}^{\text{total}}$ converge to unity as
one approaches the BZ center. This implies that if the TA or LA phonon
is less diffusely scattered by the GB, its transmission across the
GB increases correspondingly. A similar trend is observed for TA and
LA phonons in zigzag-edge graphene.

Conversely, Fig.~\ref{fig:ModeResolvedTransmissionTotalSpecularity}(d)
shows that the ZA phonon $P_{n}^{\text{total}}$ decreases in the
long-wavelength limit, consistent with the trend observed for the
ZA phonon $\mathcal{T}_{n}$ in Fig.~\ref{fig:ModeResolvedTransmissionTotalSpecularity}(a)
and confirming previous observations~\cite{STan:Carbon13_Effect,EHelgee:PRB14_Scattering,JXia:JAP17_Effect}
that long-wavelength ZA phonons are more easily scattered by the GB.
In addition, we also observe that total specularity for ZA phonons
{[}Figs.~\ref{fig:ModeResolvedTransmissionTotalSpecularity}(a) and
(d){]} is significantly smaller than unity even for ZA phonon modes
that have $\mathcal{T}_{n}\sim1$ {[}Figs.~\ref{fig:ModeResolvedTransmissionTotalSpecularity}(a)
and (d){]}. This suggests that \emph{foward} scattering for the transmitted
ZA phonons is diffuse, a phenomenon which may be due to the low-frequency
ZA phonon dispersion anisotropy in armchair-edge graphene and the
decrease in their group velocities at low frequencies.

\section{Difference between transmission and reflection specularity distribution}
To
characterize how diffusely scattered the transmitted phonons are,
we plot the transmission specularity $P_{n}^{+}$ distribution in
Figs.~\ref{fig:ModeResolvedTransmissionTotalSpecularity}(g) to (i).
Like in Figs.~\ref{fig:ModeResolvedTransmissionTotalSpecularity}(d)
to (f), the $P_{n}^{+}$ distributions are near-isotropic for the
ZA, TA and LA phonons. Nonetheless, there is a striking difference
between the ZA phonon $P_{n}^{+}$ distribution and the TA/LA phonon
$P_{n}^{+}$ distributions: as we approach the BZ center, $P_{n}^{+}$
converges to unity for the TA/LA phonons but not for the ZA phonons.
This implies that as the phonon momentum $k$ decreases, the transmitted
TA and LA phonons tend to be more specularly forward-scattered, i.e.,
the transmitted phonon energy is more concentrated into a single outgoing
phonon channel on the other side of the GB. On the other hand, Fig.~\ref{fig:ModeResolvedTransmissionTotalSpecularity}(g)
shows that the transmitted ZA phonons are partially diffusive even
in the $\omega\rightarrow0$ limit. This implies that at low frequencies,
the transmitted phonon energy is spread over multiple outgoing phonon
channels across the GB, a phenomenon also observed in Ref.~\cite{EHelgee:PRB14_Scattering}.
We also notice that the ZA, TA and LA phonons share a similarity where
$P_{n}^{+}$ generally decreases as we approach the BZ edges and corners,
indicating that shorter wavelength phonons are more diffusely scattered
by the GB. In particular, the overall $P_{n}^{+}$ distribution is
more similar to the $\mathcal{T}_{n}$ distribution than the $P_{n}^{\text{total}}$
distribution is.

For comparison with $P_{n}^{+}$ in Figs.\ref{fig:ModeResolvedTransmissionTotalSpecularity}(g)
to (i), we plot the reflection specularity $P_{n}^{-}$ distribution
in Figs.~\ref{fig:ModeResolvedTransmissionTotalSpecularity}(j) to
(l). We observe that $P_{n}^{-}<P_{n}^{+}$ as we approach the BZ
center ($\Gamma$ point) with the difference most pronounced for LA
phonons, confirming Li and McGaughey's hypothesis. On the other hand,
for the transversely polarized TA and ZA phonons, we find that $P_{n}^{-}>P_{n}^{+}$
as we approach the BZ corners ($K$ points), possibly due their inverted
Dirac cone-like phonon dispersion at the $K$ points as seen in Fig.~\ref{fig:GBStructurePhononDispersion}(b).
The absence of this behavior for the LA phonons suggests that the
phonon dispersion plays a critical role in specularity, a phenomenon
which needs to be further investigated. 

\begin{figure*}

\begin{centering}
\includegraphics[scale=0.23]{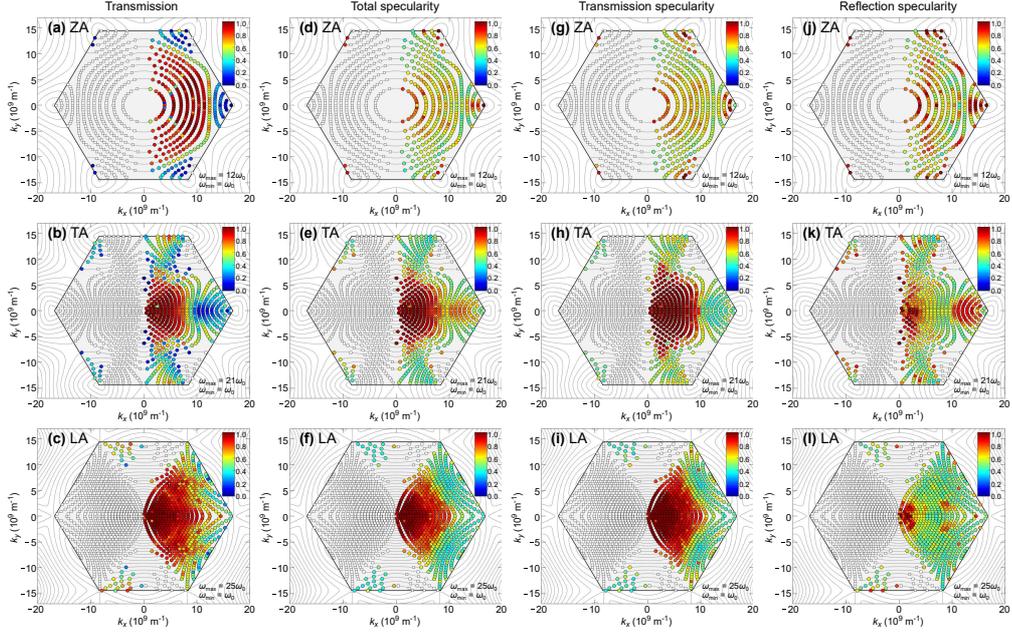}
\par\end{centering}
\caption{Mode-resolved comparison of \textbf{(a-c)} $\mathcal{T}_{n}(\omega)$,
\textbf{(d-f)} $P_{n}^{\text{total}}(\omega)$, \textbf{(g-i)} $P_{n}^{+}(\omega)$,
and \textbf{(j-l)} $P_{n}^{-}(\omega)$ for ZA, TA and LA phonons
in armchair-edge graphene at each frequency $\omega$ point over the
frequency range $\omega=m\omega_{0}$ for $m=$ 1 to 25 and $\omega_{0}=10^{13}$
rad/s ($6.58$ meV) for a (32,32)$|$(56,0) graphene GB constructed from
only GB-II subunits. The isofrequency contours at each $\omega$ are
shown using solid gray lines. The modes in the incoming phonon flux
are represented by filled circles, colored according to their numerical
value as indicated in the color bars, while the modes in the outgoing
flux are represented by hollow squares.}

\label{fig:ModeResolvedTransmissionTotalSpecularity_GB2}
\end{figure*}

\section{Role of disorder}
To analyze the role of disorder
in phonon scattering, we compute and plot $\mathcal{T}_{n}$, $P_{n}^{\text{total}}$,
$P_{n}^{+}$ and $P_{n}^{-}$ in Fig.~\ref{fig:ModeResolvedTransmissionTotalSpecularity_GB2}
for a (32,32)$|$(56,0) graphene GB constructed from only GB-II subunits.
Physically, this GB configuration is a regular array of GB-II subunits
and less disordered than the earlier GB models comprising a random
mix of GB-II and GB-III subunits. In the following discussion, we
refer to it as the ordered GB model and the earlier GB models collectively
as the disordered GB model. In terms of phonon transmission ($\mathcal{T}_{n}$),
we do not observe any significant difference between the ordered and
disordered GB models except in Fig.~\ref{fig:ModeResolvedTransmissionTotalSpecularity_GB2}(a)
and Fig.~\ref{fig:ModeResolvedTransmissionTotalSpecularity}(a) where
the low-frequency ZA phonons at $\omega=10^{13}$ rad/s are more easily
transmitted in the ordered GB model. This suggests that the reduction
in $\mathcal{T}_{n}$ for low-frequency ZA phonons in the disordered
GB model is caused by the increased scattering with disorder. We find
further support for this explanation by comparing $P_{n}^{\text{total}}$,
$P_{n}^{+}$ and $P_{n}^{-}$ for the ZA phonons in the ordered and
disordered GB models. Figures~\ref{fig:ModeResolvedTransmissionTotalSpecularity_GB2}(d),
(g) and (j) show that $P_{n}^{\text{total}}$, $P_{n}^{+}$ and $P_{n}^{-}$
are higher for the ordered GB model compared to the results in Figs.~\ref{fig:ModeResolvedTransmissionTotalSpecularity}(d),
(g) and (j). We also find that the $P_{n}^{+}$ and $P_{n}^{-}$ distribution
for ZA phonons are comparable for the ordered GB unlike the $P_{n}^{+}>P_{n}^{-}$
trend for the disordered GB. For the TA and LA phonons, the difference
between the $P_{n}^{+}$ and $P_{n}^{-}$ distribution is less pronounced
in the ordered GB especially at how frequencies near the center of
the BZ where $P_{n}^{+}$ and $P_{n}^{-}$ are comparable and close
to unity. Away from the BZ center, we find that $P_{n}^{+}>P_{n}^{-}$
in the ordered GB model although it is less pronounced than the disordered
GB model. This is probably because the ordered GB model does not have
a perfectly flat interface as it is impossible to form a commensurate
boundary between armchair-edge and zigzag-edge graphene.

\section{Summary and conclusion}
We have investigated the
difference in specularity between phonon reflection and transmission
at a graphene GB. Using the atomistic $S$-matrix method, we analyze
the phonon transmission probability $\mathcal{T}_{n}$ and its relationship
to the total, transmission and reflection specularity parameters ($P_{n}^{\text{total}}$,
$P_{n}^{+}$ and $P_{n}^{-}$) for the LA, TA and ZA phonons. We find
that the $\mathcal{T}_{n}$ and $P_{n}^{\text{+}}$ distributions
are highly similar over the entire Brillouin Zone. 
There is a striking
difference between $P_{n}^{+}$ and $P_{n}^{-}$, indicating that
the degrees of randomization for the transmitted and reflected phonons
are dissimilar. We confirm Li and McGaughey's hypothesis that $P_{n}^{+}>P_{n}^{-}$
especially nearer the $\Gamma$ point, with the difference most pronounced
for LA phonons, and find that $P_{n}^{+}<P_{n}^{-}$ nearer the $K$
point for TA and ZA phonons, possibly because of their inverted Dirac
cone-like phonon dispersion. Our results show that the phonon dispersion
plays a critical role in transmission and specularity.

\acknowledgments
We acknowledge financial support from a grant from the Science and
Engineering Research Council (Grant No. 152-70-00017) and the Agency
for Science, Technology, and Research (A{*}STAR), Singapore.

\bibliographystyle{eplbib}
\bibliography{PaperReferences,LocalReferences}

\end{document}


\title{Supplemental material: Specular transmission and diffuse reflection
in phonon scattering at grain boundary}
\author{Zhun-Yong Ong}
\affiliation{Institute of High Performance Computing, A{*}STAR, Singapore 138632,
Singapore}
\email{ongzy@ihpc.a-star.edu.sg}

\maketitle
\tableofcontents{}

\renewcommand{\thesection}{S\arabic{section}}    
\renewcommand{\thetable}{S\arabic{table}}    
\renewcommand{\thefigure}{S\arabic{figure}}
\renewcommand{\theequation}{S\arabic{equation}}

\section{Grain boundary structure generation and optimization\label{sec:GBStructureGeneration}}

\begin{figure}
\begin{centering}
\includegraphics[scale=0.5]{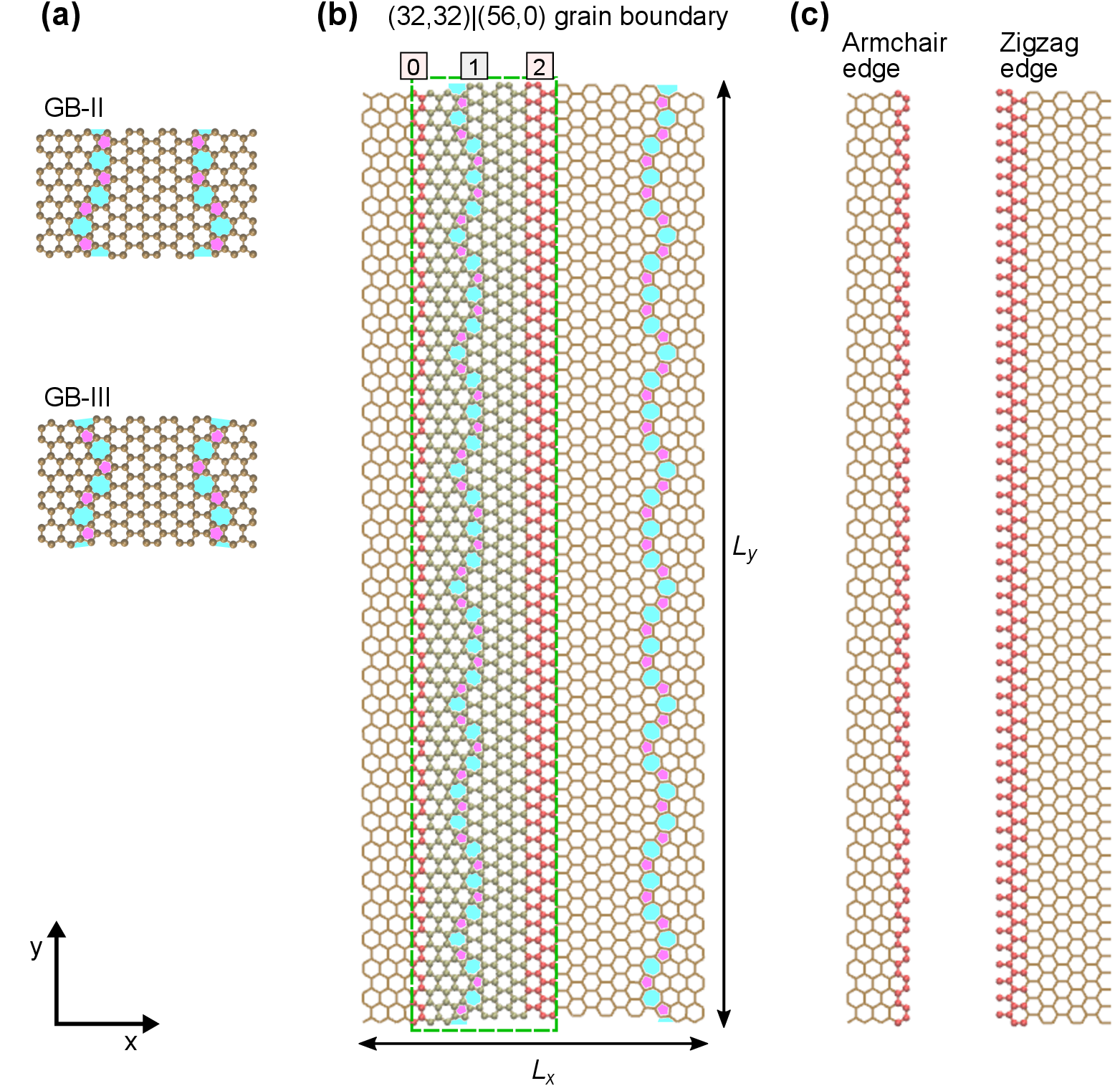}
\par\end{centering}
\caption{\textbf{(a)} Atomic structure of the original GB-II and GB-III supercells
with (4,4)/(7,0) graphene grain boundaries, taken from Ref.~\citep{GSchusteritsch:PRB14_Predicting}.
\textbf{(b)} Example of a (32,32)|(56,0) grain boundary constructed
from GB-II and GB-III. The scattering region is enclosed in green
dashed lines and divided into three subregions, 0 to 2, with 0 and
2 corresponding to the edges coupled to the bulk. \textbf{(c)} Atomic
structure of bulk armchair- and zigzag-edge graphene, comprising layers
arrayed in the $x$-direction, with periodic boundary conditions in
the $x$-direction. The red atoms correspond to one layer.}

\label{fig:AtomicPositions}
\end{figure}

The generation and optimization of the graphene grain boundary (GB)
structures used in our simulation are identical to those in Ref.~\citep{ZYOng:PRB20_Structure}.
The original 186-atom GB-II and GB-III supercells from Ref.~\citep{GSchusteritsch:PRB14_Predicting},
which are used to construct the (32,32)|(56,0) GB interfaces, are
shown in Fig.~\ref{fig:AtomicPositions}(a). Each supercell has two
(4,4)|(7,0) grain boundaries consisting of a continuous undulating
line of pentagon-heptagon defect pairs that define the interface between
bulk armchair and zigzag-edge graphene. By `zigzag' or `armchair'
edge, we refer to the orientation of the carbon atoms in the $y$-direction
parallel to the GB. 

To generate the larger GB structure in Fig.~\ref{fig:AtomicPositions}(b),
we first expand GB-II and GB-III in the $x$ direction by increasing
the size of the bulk armchair and zigzag-edge graphene region between
the two interfaces, to reduce the distortion of the bulk lattice region.
We then line $n_{\text{GB}}=8$ of the expanded GB-II's and GB-III's
up in the $y$-direction to create the (32,32)|(56,0) GB structure
in Fig.~\ref{fig:AtomicPositions}(b), with the exact configuration
of the (32,32)|(56,0) GB depending on the number of GB-II's and GB-III's
used and the order in which they are arranged.

We model the interatomic interaction using the Tersoff potential~\citep{JTersoff:PRL88_Empirical}
with parameters from Ref.~\citep{LLindsay:PRB10_Optimized}. In bulk
graphene, the C-C bond length of $1.44\text{Å}$ minimizes the total
potential energy. Because the zigzag edge is slightly wider than the
armchair edge by about one percent, we use this value to fix the width
of the GB as $L_{y}=139.55\lyxmathsym{\AA}$ as this results in the
minimized average C-C bond length of $1.44\text{Å}$ and $1.45\text{Å}$
in bulk zigzag-edge and armchair-edge graphene, respectively. The
slightly longer C-C bond length in armchair-edge graphene is due to
the one-percent tensile strain in the $y$-direction. We use the code
GULP~\citep{JGale:MolSim03_gulp} to optimize the atomic positions.
The supercell in Fig.~\ref{fig:AtomicPositions}(b) is allowed to
relax only in the $x$ direction, with periodic boundary conditions
are imposed in the $x$ and $y$direction, until the optimized value
of $L_{x}$, the length of the supercell in the $x$-direction, is
obtained. After minimization, the C-C bond lengths in the scattering
region bounded by dashed lines in Fig.~\ref{fig:AtomicPositions}(b)
vary between $1.39\text{Å}$ and $1.49\text{Å}$.

After minimization, the force constants corresponding to the atoms
in the scattering region are computed in GULP and then postprocessed
to yield the interatomic force constant (IFC) matrix describing the
scattering region, 
\begin{equation}
\boldsymbol{K}_{\text{C}}=\left(\begin{array}{ccc}
\boldsymbol{K}_{00} & \boldsymbol{K}_{01}\\
\boldsymbol{K}_{10} & \boldsymbol{K}_{11} & \boldsymbol{K}_{12}\\
 & \boldsymbol{K}_{21} & \boldsymbol{K}_{22}
\end{array}\right)\label{eq:CentralMatrix}
\end{equation}
where the submatrix $\boldsymbol{K}_{nm}$ describes the coupling
between subregions $n$ and $m$ in Fig.~\ref{fig:AtomicPositions}(b).
Subregions 0 and 2 represent the bulk-like edges of the scattering
region and connect the scattering region to bulk armchair- and zigzag-edge
graphene while subregion 1 corresponds to the center of the scattering
region. We also extract the matrices $\boldsymbol{K}_{\text{L}}^{00}$
and $\boldsymbol{K}_{\text{L}}^{01}$ ($\boldsymbol{K}_{\text{R}}^{00}$
and $\boldsymbol{K}_{\text{R}}^{01}$) which describe the coupling
within and between layers, as shown in Fig.~\ref{fig:AtomicPositions}(c),
in armchair-edge (zigzag-edge) graphene. The block-tridiagonal IFC
matrix for the entire scattering system is 
\begin{equation}
\mathbf{K}=\begin{pmatrix}\begin{array}{c|c|c}
\begin{array}{cc}
\ddots & \ddots\\
\ddots & \boldsymbol{K}_{\text{L}}^{00}
\end{array} & \begin{array}{ccc}
\vphantom{\ddots} & \hphantom{\boldsymbol{K}_{11}} & \hphantom{\boldsymbol{K}_{22}}\\
\boldsymbol{K}_{\text{L}}^{01} & \vphantom{\ddots} & \vphantom{\ddots}
\end{array}\\
\hline \begin{array}{cc}
\hphantom{\ddots} & (\boldsymbol{K}_{\text{L}}^{01})^{\dagger}\\
\\
\\
\end{array} & \begin{array}{ccc}
\boldsymbol{K}_{00} & \boldsymbol{K}_{01}\\
\boldsymbol{K}_{10} & \boldsymbol{K}_{11} & \boldsymbol{K}_{12}\\
 & \boldsymbol{K}_{21} & \boldsymbol{K}_{22}
\end{array} & \begin{array}{cc}
\vphantom{\boldsymbol{K}_{\text{R}}^{00}}\\
\vphantom{\boldsymbol{K}_{\text{R}}^{00}}\\
\boldsymbol{K}_{\text{R}}^{01} & \hphantom{\ddots}
\end{array}\\
\hline  & \begin{array}{ccc}
\hphantom{\boldsymbol{K}_{\text{L}}^{00}} & \hphantom{\boldsymbol{K}_{\text{L}}^{00}} & (\boldsymbol{K}_{\text{R}}^{01})^{\dagger}\\
 &  & \vphantom{\ddots}
\end{array} & \begin{array}{cc}
\boldsymbol{K}_{\text{R}}^{00} & \ddots\\
\ddots & \ddots
\end{array}
\end{array}\end{pmatrix}\ ,\label{eq:SystemFCMatrix}
\end{equation}
which we rewrite as 
\[
\mathbf{K}=\left(\begin{array}{ccccc}
\ddots & \ddots\\
\ddots & \boldsymbol{K}_{\text{L}}^{00} & \boldsymbol{K}_{\text{CL}}^{\dagger}\\
 & \boldsymbol{K}_{\text{CL}} & \boldsymbol{K}_{\text{C}} & \boldsymbol{K}_{\text{CR}}\\
 &  & \boldsymbol{K}_{\text{CR}}^{\dagger} & \boldsymbol{K}_{\text{R}}^{00} & \ddots\\
 &  &  & \ddots & \ddots
\end{array}\right)
\]
where $\boldsymbol{K}_{\text{CL}}$ ($\boldsymbol{K}_{\text{CR}}$
) represents the coupling between the scattering region and the left
(right) lead and is derived from $\boldsymbol{K}_{\text{L}}^{01}$
($\boldsymbol{K}_{\text{R}}^{01}$). To satisfy the acoustic sum rule
required from translational invariance, we recalculate the diagonal
elements of $\mathbf{K}$ to satisfy $[\mathbf{K}]_{ii}=-\sum_{j\neq i}[\mathbf{K}]_{ij}$.
Finally, the mass-normalized force-constant matrices $\boldsymbol{H}_{\text{L}}^{00}$,
$\boldsymbol{H}_{\text{L}}^{01}$, $\boldsymbol{H}_{\text{R}}^{00}$,
$\boldsymbol{H}_{\text{R}}^{01}$, $\boldsymbol{H}_{\text{CL}}$,
$\boldsymbol{H}_{\text{C}}$ and $\boldsymbol{H}_{\text{CR}}$ are
obtained from the IFC matrices by dividing each matrix by $m$, the
atomic mass of carbon.

\section{S-matrix calculation methodology}

In this section, we give the details of the $S$ matrix calculations
which are based on the $S$-matrix method described in Ref.~\citep{ZYOng:PRB18_Atomistic}.
The $S$ matrix is a frequency-dependent matrix in which the matrix
elements are the scattering amplitudes between the incoming and outgoing
scattering channels which correspond to the bulk phonon modes in armchair
and zigzag-edge graphene. 

The Green's function of the scattering region in Fig.~\ref{fig:AtomicPositions}(b)
is given by 
\[
\boldsymbol{G}_{\text{C}}^{\text{ret}}(\omega)=[(\omega^{2}+i0^{+})\boldsymbol{I}-\boldsymbol{H}_{\text{C}}-\boldsymbol{\Sigma}_{\text{L}}(\omega)-\boldsymbol{\Sigma}_{\text{R}}(\omega)]^{-1}
\]
 where $\boldsymbol{\Sigma}_{\text{L}}(\omega)=\boldsymbol{H}_{\text{CL}}\boldsymbol{g}_{\text{L},-}^{\text{ret}}(\omega)\boldsymbol{H}_{\text{CL}}^{\dagger}$
and $\boldsymbol{\Sigma}_{\text{R}}(\omega)=\boldsymbol{H}_{\text{CR}}\boldsymbol{g}_{\text{R},+}^{\text{ret}}(\omega)\boldsymbol{H}_{\text{CR}}^{\dagger}$
are the self-energies associated with the left (i.e. bulk armchair-edge
graphene) and right (i.e. bulk zigzag-edge graphene) leads, respectively,
with the left and right-lead \emph{retarded} surface Green's functions
given by \begin{subequations}
\begin{equation}
\boldsymbol{g}_{\text{L},-}^{\text{ret}}(\omega)=[(\omega^{2}+i0^{+})\boldsymbol{I}_{\text{L}}-\boldsymbol{H}_{\text{L}}^{00}-(\boldsymbol{H}_{\text{L}}^{01})^{\dagger}\boldsymbol{g}_{\text{L},-}^{\text{ret}}(\omega)\boldsymbol{H}_{\text{L}}^{01}]^{-1}\label{eq:LeftRetardedSurfaceGF_minus}
\end{equation}
\begin{equation}
\boldsymbol{g}_{\text{R},+}^{\text{ret}}(\omega)=[(\omega^{2}+i0^{+})\boldsymbol{I}_{\text{R}}-\boldsymbol{H}_{\text{R}}^{00}-\boldsymbol{H}_{\text{R}}^{01}\boldsymbol{g}_{\text{R},+}^{\text{ret}}(\omega)(\boldsymbol{H}_{\text{R}}^{01})^{\dagger}]^{-1}\ .\label{eq:RightRetardedSurfaceGF_plus}
\end{equation}
$\boldsymbol{I}$, $\boldsymbol{I}_{\text{L}}$ and $\boldsymbol{I}_{\text{R}}$
are identity matrices of the same size as $\boldsymbol{H}_{\text{C}}$,
$\boldsymbol{H}_{\text{L}}^{00}$ and $\boldsymbol{H}_{\text{R}}^{00}$,
respectively. In addition, we also define 
\begin{equation}
\boldsymbol{g}_{\text{L},+}^{\text{ret}}(\omega)=[(\omega^{2}+i0^{+})\boldsymbol{I}_{\text{L}}-\boldsymbol{H}_{\text{L}}^{00}-\boldsymbol{H}_{\text{L}}^{01}\boldsymbol{g}_{\text{L},+}^{\text{ret}}(\omega)(\boldsymbol{H}_{\text{L}}^{01})^{\dagger}]^{-1}\ ,\label{LeftRetardedSurfaceGF_plus}
\end{equation}
\begin{equation}
\boldsymbol{g}_{\text{R},-}^{\text{ret}}(\omega)=[(\omega^{2}+i0^{+})\boldsymbol{I}_{\text{R}}-\boldsymbol{H}_{\text{R}}^{00}-(\boldsymbol{H}_{\text{R}}^{01})^{\dagger}\boldsymbol{g}_{\text{R},-}^{\text{ret}}(\omega)\boldsymbol{H}_{\text{R}}^{01}]^{-1}\ ,\label{eq:RightRetardedSurfaceGF_minus}
\end{equation}
\label{subeq:RetardedSurfaceGFs}\end{subequations} $\boldsymbol{g}_{\text{L},-}^{\text{adv}}(\omega)=\boldsymbol{g}_{\text{L},-}^{\text{ret}}(\omega)^{\dagger}$
and $\boldsymbol{g}_{\text{R},+}^{\text{adv}}(\omega)=\boldsymbol{g}_{\text{R},+}^{\text{ret}}(\omega)^{\dagger}$.
The retarded surface Green's functions in Eq.~(\ref{subeq:RetardedSurfaceGFs})
can be computed using established numerical methods~\citep{JSWang:EPJB08_Quantum}
and exploiting the transverse translational symmetry~\citep{ZYOng:PRB18_Atomistic}.

Given Eq.~(\ref{subeq:RetardedSurfaceGFs}), we define the Bloch
matrices ($\boldsymbol{F}_{\text{L}}^{\text{ret/adv}}(-)$ and $\boldsymbol{F}_{\text{R}}^{\text{ret/adv}}(+)$):
\[
\boldsymbol{F}_{\text{L}}^{\text{ret/adv}}(-)^{-1}=\boldsymbol{g}_{\text{L},-}^{\text{ret/adv}}\boldsymbol{H}_{\text{L}}^{01}
\]
\[
\boldsymbol{F}_{\text{R}}^{\text{ret/adv}}(+)=\boldsymbol{g}_{\text{R},+}^{\text{ret/adv}}\boldsymbol{H}_{\text{R}}^{10}
\]
and their associated eigenmode matrices ($\boldsymbol{U}_{\text{L}}^{\text{ret/adv}}(-)$
and $\boldsymbol{U}_{\text{R}}^{\text{ret/adv}}(+)$): \begin{subequations}
\begin{equation}
\boldsymbol{F}_{\text{L}}^{\text{ret/adv}}(-)^{-1}\boldsymbol{U}_{\text{L}}^{\text{ret/adv}}(-)=\boldsymbol{U}_{\text{L}}^{\text{ret/adv}}(-)\boldsymbol{\Lambda}_{\text{L}}^{\text{ret/adv}}(-)^{-1}\label{eq:LeftBlochMatrices}
\end{equation}
\begin{equation}
\boldsymbol{F}_{\text{R}}^{\text{ret/adv}}(+)\boldsymbol{U}_{\text{R}}^{\text{ret/adv}}(+)=\boldsymbol{U}_{\text{R}}^{\text{ret/adv}}(+)\boldsymbol{\Lambda}_{\text{R}}^{\text{ret/adv}}(+)\label{eq:RightBlochMatrices}
\end{equation}
\label{subeq:EigenmodeMatrices}\end{subequations}along with the
eigenvelocity matrices ($\boldsymbol{V}_{\text{L}}^{\text{ret/adv}}(-)$
and $\boldsymbol{V}_{\text{R}}^{\text{ret/adv}}(+)$): \begin{subequations}
\begin{equation}
\boldsymbol{V}_{\text{L}}^{\text{ret/adv}}(-)=-\frac{ia_{\text{L}}}{2\omega}[\boldsymbol{U}_{\text{L}}^{\text{ret/adv}}(-)]^{\dagger}\boldsymbol{H}_{\text{L}}^{10}[\boldsymbol{g}_{\text{L},-}^{\text{ret/adv}}-(\boldsymbol{g}_{\text{L},-}^{\text{ret/adv}})^{\dagger}]\boldsymbol{H}_{\text{L}}^{01}\boldsymbol{U}_{\text{L}}^{\text{ret/adv}}(-)\label{eq:LeftVelocityMatrices}
\end{equation}
\begin{equation}
\boldsymbol{V}_{\text{R}}^{\text{ret/adv}}(+)=\frac{ia_{\text{R}}}{2\omega}[\boldsymbol{U}_{\text{R}}^{\text{ret/adv}}(+)]^{\dagger}\boldsymbol{H}_{\text{R}}^{01}[\boldsymbol{g}_{\text{R},+}^{\text{ret/adv}}-(\boldsymbol{g}_{\text{R},+}^{\text{ret/adv}})^{\dagger}]\boldsymbol{H}_{\text{R}}^{10}\boldsymbol{U}_{\text{R}}^{\text{ret/adv}}(+)\label{eq:RightVelocityMatrices}
\end{equation}
\label{subeq:EigenvelocityMatrices}\end{subequations} Dropping the
subscripts and superscripts, we can write the eigenmode matrices in
Eq.~(\ref{subeq:EigenmodeMatrices}) as 
\begin{equation}
\boldsymbol{U}=(\mathbf{u}(\boldsymbol{k}_{1}),\mathbf{u}(\boldsymbol{k}_{2}),\ldots,\mathbf{u}(\boldsymbol{k}_{N}))\label{eq:EigenmodeMatrix}
\end{equation}
where $\mathbf{u}(\boldsymbol{k}_{n})$ is a column vector corresponding
to the phonon eigenmode with wave vector $\boldsymbol{k}_{n}$ which
is real for bulk phonon modes and complex for evanescent modes. The
wave vector $\boldsymbol{k}_{n}$ is determined using the zone-unfolding
procedure described in Ref.~\citep{ZYOng:PRB18_Atomistic} and is
also identified according to its polarization/branch. We may regard
each column vector of $\boldsymbol{U}_{\text{L}}^{\text{adv}}(-)$
{[}$\boldsymbol{U}_{\text{L}}^{\text{ret}}(-)${]} as an incoming
(outgoing) left-lead phonon mode describing the atomic displacements
in the bulk layer immediately left of subregion 0 in Fig.~\ref{fig:AtomicPositions}(b)
and likewise each column vector of $\boldsymbol{U}_{\text{R}}^{\text{adv}}(+)$
{[}$\boldsymbol{U}_{\text{R}}^{\text{ret}}(+)${]} as an incoming
(outgoing) right-lead phonon mode describing the atomic displacements
in the bulk layer immediately right of subregion 2 in Fig.~\ref{fig:AtomicPositions}(a).
Although there are $N$ column vectors on the right hand side of Eq.~(\ref{eq:EigenmodeMatrix})
at each frequency $\omega$, only a subset of them correspond to bulk
phonon modes, with the number of bulk phonon modes varying with the
frequency.

In the following discussion, we reserve the symbols $\boldsymbol{p}_{n}$
and $\boldsymbol{p}_{n}^{\prime}$ (where $n=1,\ldots,N_{\text{L}}(\omega)$
and $N_{\text{L}}(\omega)$ is the number of outgoing or incoming
phonon modes at frequency $\omega$) for the wave vectors of the outgoing
and incoming \emph{left}-lead bulk phonons, respectively, and $\boldsymbol{q}_{m}$
and $\boldsymbol{q}_{m}^{\prime}$ (where $m=1,\ldots,N_{\text{R}}(\omega)$
and $N_{\text{R}}(\omega)$ is the number of outgoing or incoming
\emph{right}-lead phonon modes at frequency $\omega$) for the wave
vectors of the outgoing and incoming right-lead bulk phonons, respectively.
The incoming bulk phonon modes are marked with a $\prime$ superscript
while the outgoing ones are not. At each frequency $\omega$, we have
a set of outgoing left-lead phonons associated with the wave vectors
$\mathcal{U}_{\text{L}}^{\text{ret}}(\omega)=\{\boldsymbol{p}_{1},\ldots,\boldsymbol{p}_{N_{\text{L}}(\omega)}\}$.
The number of elements as well as the composition of this set depends
on $\omega$. So, at another frequency $\omega^{\prime}\neq\omega$,
we have a different set of left-lead phonons $\mathcal{U}_{\text{L}}^{\text{ret}}(\omega^{\prime})=\{\boldsymbol{p}_{1},\ldots,\boldsymbol{p}_{N_{\text{L}}(\omega^{\prime})}\}$
and its wave vectors are usually different to those of $\mathcal{U}(\omega)$
unless the phonons belong to different polarization branches. Therefore,
at each frequency $\omega$, we have two sets of wave vectors $\mathcal{U}_{\text{L}}^{\text{adv}}(\omega)=\{\boldsymbol{p}_{1}^{\prime},\ldots,\boldsymbol{p}_{N_{\text{L}}(\omega)}^{\prime}\}$
and $\mathcal{U}_{\text{R}}^{\text{adv}}(\omega)=\{\boldsymbol{q}_{1}^{\prime},\ldots,\boldsymbol{q}_{N_{\text{R}}(\omega)}^{\prime}\}$
associated with the incoming left and right-lead phonons, respectively,
and another two sets of wave vectors $\mathcal{U}_{\text{L}}^{\text{ret}}(\omega)=\{\boldsymbol{p}_{1},\ldots,\boldsymbol{p}_{N_{\text{L}}(\omega)}\}$
and $\mathcal{U}_{\text{R}}^{\text{ret}}(\omega)=\{\boldsymbol{q}_{1},\ldots,\boldsymbol{q}_{N_{\text{R}}(\omega)}\}$
associated with the outgoing left and right-lead phonons, respectively.

Given Eqs.~(\ref{subeq:EigenmodeMatrices}) and (\ref{subeq:EigenvelocityMatrices}),
the formulas for the transmission and reflection matrices are:~\citep{ZYOng:PRB18_Atomistic}
\begin{subequations}
\begin{equation}
\boldsymbol{t}_{\text{RL}}(\omega)=\frac{2i\omega}{\sqrt{a_{\text{R}}a_{\text{L}}}}[\boldsymbol{V}_{\text{R}}^{\text{ret}}(+)]^{1/2}[\boldsymbol{U}_{\text{R}}^{\text{ret}}(+)]^{-1}\boldsymbol{G}_{\text{RL}}^{\text{ret}}[\boldsymbol{U}_{\text{L}}^{\text{adv}}(-)^{\dagger}]^{-1}[\boldsymbol{V}_{\text{L}}^{\text{adv}}(-)]^{1/2}\label{eq:Left2RightTransmissionMatrices}
\end{equation}
\begin{equation}
\boldsymbol{t}_{\text{LR}}(\omega)=\frac{2i\omega}{\sqrt{a_{\text{L}}a_{\text{R}}}}[\boldsymbol{V}_{\text{L}}^{\text{ret}}(-)]^{1/2}[\boldsymbol{U}_{\text{L}}^{\text{ret}}(-)]^{-1}\boldsymbol{G}_{\text{LR}}^{\text{ret}}[\boldsymbol{U}_{\text{R}}^{\text{adv}}(+)^{\dagger}]^{-1}[\boldsymbol{V}_{\text{R}}^{\text{adv}}(+)]^{1/2}\label{eq:Right2LeftTransmissionMatrices}
\end{equation}
\begin{equation}
\boldsymbol{r}_{\text{LL}}(\omega)=\frac{2i\omega}{a_{\text{L}}}[\boldsymbol{V}_{\text{L}}^{\text{ret}}(-)]^{1/2}[\boldsymbol{U}_{\text{L}}^{\text{ret}}(-)]^{-1}(\boldsymbol{G}_{\text{L}}^{\text{ret}}-\boldsymbol{Q}_{\text{L}}^{-1})[\boldsymbol{U}_{\text{L}}^{\text{adv}}(-)^{\dagger}]^{-1}[\boldsymbol{V}_{\text{L}}^{\text{adv}}(-)]^{1/2}\label{eq:LeftReflectionMatrices}
\end{equation}
\begin{equation}
\boldsymbol{r}_{\text{RR}}(\omega)=\frac{2i\omega}{a_{\text{R}}}[\boldsymbol{V}_{\text{R}}^{\text{ret}}(+)]^{1/2}[\boldsymbol{U}_{\text{R}}^{\text{ret}}(+)]^{-1}(\boldsymbol{G}_{\text{R}}^{\text{ret}}-\boldsymbol{Q}_{\text{R}}^{-1})[\boldsymbol{U}_{\text{R}}^{\text{adv}}(+)^{\dagger}]^{-1}[\boldsymbol{V}_{\text{R}}^{\text{adv}}(+)]^{1/2}\label{eq:RightReflectionMatrices}
\end{equation}
\label{subeq:TransmissionReflectionMatrices}\end{subequations}where
\begin{subequations}
\[
\boldsymbol{G}_{\text{RL}}^{\text{ret}}(\omega)=\boldsymbol{g}_{\text{R},+}^{\text{ret}}(\omega)\boldsymbol{H}_{\text{RC}}\boldsymbol{G}_{\text{C}}^{\text{ret}}(\omega)\boldsymbol{H}_{\text{CL}}\boldsymbol{g}_{\text{L},-}^{\text{ret}}(\omega)
\]
\[
\boldsymbol{G}_{\text{LR}}^{\text{ret}}(\omega)=\boldsymbol{g}_{\text{L},-}^{\text{ret}}(\omega)\boldsymbol{H}_{\text{LC}}\boldsymbol{G}_{\text{C}}^{\text{ret}}(\omega)\boldsymbol{H}_{\text{CR}}\boldsymbol{g}_{\text{R},+}^{\text{ret}}(\omega)
\]
\[
\boldsymbol{G}_{\text{L}}^{\text{ret}}(\omega)=\boldsymbol{g}_{\text{L},-}^{\text{ret}}(\omega)+\boldsymbol{g}_{\text{L},-}^{\text{ret}}(\omega)\boldsymbol{H}_{\text{LC}}\boldsymbol{G}_{\text{C}}^{\text{ret}}(\omega)\boldsymbol{H}_{\text{CL}}\boldsymbol{g}_{\text{L},-}^{\text{ret}}(\omega)
\]
\[
\boldsymbol{G}_{\text{R}}^{\text{ret}}(\omega)=\boldsymbol{g}_{\text{R},+}^{\text{ret}}(\omega)+\boldsymbol{g}_{\text{R},+}^{\text{ret}}(\omega)\boldsymbol{H}_{\text{RC}}\boldsymbol{G}_{\text{C}}^{\text{ret}}(\omega)\boldsymbol{H}_{\text{CR}}\boldsymbol{g}_{\text{R},+}^{\text{ret}}(\omega)
\]
\[
\boldsymbol{Q}_{\text{L}}(\omega)=(\omega^{2}+i0^{+})\boldsymbol{I}_{\text{L}}-\boldsymbol{H}_{\text{L}}^{00}-(\boldsymbol{H}_{\text{L}}^{01})^{\dagger}\boldsymbol{g}_{\text{L},-}^{\text{ret}}(\omega)\boldsymbol{H}_{\text{L}}^{01}-\boldsymbol{H}_{\text{L}}^{01}\boldsymbol{g}_{\text{L},+}^{\text{ret}}(\omega)(\boldsymbol{H}_{\text{L}}^{01})^{\dagger}
\]
\[
\boldsymbol{Q}_{\text{R}}(\omega)=(\omega^{2}+i0^{+})\boldsymbol{I}_{\text{R}}-\boldsymbol{H}_{\text{R}}^{00}-(\boldsymbol{H}_{\text{R}}^{01})^{\dagger}\boldsymbol{g}_{\text{R},-}^{\text{ret}}(\omega)\boldsymbol{H}_{\text{R}}^{01}-\boldsymbol{H}_{\text{R}}^{01}\boldsymbol{g}_{\text{R},+}^{\text{ret}}(\omega)(\boldsymbol{H}_{\text{R}}^{01})^{\dagger}
\]
\end{subequations}We obtain the reduced transmission and reflection
matrices $\overline{\boldsymbol{t}}_{\text{RL}}$, $\overline{\boldsymbol{t}}_{\text{LR}}$,
$\overline{\boldsymbol{r}}_{\text{LL}}$ and $\overline{\boldsymbol{r}}_{\text{RR}}$
from Eq.~(\ref{subeq:TransmissionReflectionMatrices}) by eliminating
the matrix columns and rows associated with the evanescent modes.
The matrix $\overline{\boldsymbol{t}}_{\text{RL}}$ is an $N_{\text{R}}\times N_{\text{L}}$
matrix of the form: 
\begin{equation}
\overline{\boldsymbol{t}}_{\text{RL}}=\left(\begin{array}{ccc}
S(\boldsymbol{q}_{1},\boldsymbol{p}_{1}^{\prime}) & \ldots & S(\boldsymbol{q}_{1},\boldsymbol{p}_{N_{\text{L}}}^{\prime})\\
\vdots & \ddots & \vdots\\
S(\boldsymbol{q}_{N_{\text{R}}},\boldsymbol{p}_{1}^{\prime}) & \ldots & S(\boldsymbol{q}_{N_{\text{R}}},\boldsymbol{p}_{N_{\text{L}}}^{\prime})
\end{array}\right)\label{eq:tRLmatrices}
\end{equation}
where $S(\boldsymbol{q}_{n},\boldsymbol{p}_{m}^{\prime})$ is the
complex scattering amplitude between the incoming left-lead $\boldsymbol{p}_{m}^{\prime}$
phonon mode and the outgoing right-lead $\boldsymbol{q}_{n}$ phonon
mode. Similarly, the remaining transmission and reflection matrices
are 
\begin{equation}
\overline{\boldsymbol{r}}_{\text{LL}}=\left(\begin{array}{ccc}
S(\boldsymbol{p}_{1},\boldsymbol{p}_{1}^{\prime}) & \ldots & S(\boldsymbol{p}_{1},\boldsymbol{p}_{N_{\text{L}}}^{\prime})\\
\vdots & \ddots & \vdots\\
S(\boldsymbol{p}_{N_{\text{L}}},\boldsymbol{p}_{1}^{\prime}) & \ldots & S(\boldsymbol{p}_{N_{\text{L}}},\boldsymbol{p}_{N_{\text{L}}}^{\prime})
\end{array}\right)\label{eq:rLLmatrices}
\end{equation}
\begin{equation}
\overline{\boldsymbol{t}}_{\text{LR}}=\left(\begin{array}{ccc}
S(\boldsymbol{p}_{1},\boldsymbol{q}_{1}^{\prime}) & \ldots & S(\boldsymbol{p}_{1},\boldsymbol{q}_{N_{\text{R}}}^{\prime})\\
\vdots & \ddots & \vdots\\
S(\boldsymbol{p}_{N_{\text{L}}},\boldsymbol{q}_{1}^{\prime}) & \ldots & S(\boldsymbol{p}_{N_{\text{L}}},\boldsymbol{q}_{N_{\text{R}}}^{\prime})
\end{array}\right)\label{eq:tLRmatrices}
\end{equation}
\begin{equation}
\overline{\boldsymbol{r}}_{\text{RR}}=\left(\begin{array}{ccc}
S(\boldsymbol{q}_{1},\boldsymbol{q}_{1}^{\prime}) & \ldots & S(\boldsymbol{q}_{1},\boldsymbol{q}_{N_{\text{L}}}^{\prime})\\
\vdots & \ddots & \vdots\\
S(\boldsymbol{q}_{N_{\text{R}}},\boldsymbol{q}_{1}^{\prime}) & \ldots & S(\boldsymbol{q}_{N_{\text{L}}},\boldsymbol{q}_{N_{\text{L}}}^{\prime})
\end{array}\right)\label{eq:rRRmatrices}
\end{equation}
Equations (\ref{eq:tRLmatrices}) to (\ref{eq:rRRmatrices}) are combined
to yield the overall $N_{\text{S}}\times N_{\text{S}}$ scattering
matrix: 
\begin{equation}
\boldsymbol{S}(\omega)=\begin{pmatrix}\begin{array}{cc}
\overline{\boldsymbol{r}}_{\text{LL}} & \overline{\boldsymbol{t}}_{\text{LR}}\\
\overline{\boldsymbol{t}}_{\text{RL}} & \overline{\boldsymbol{r}}_{\text{RR}}
\end{array}\end{pmatrix}=\begin{pmatrix}\begin{array}{c|c}
\begin{array}{ccc}
S(\boldsymbol{p}_{1},\boldsymbol{p}_{1}^{\prime}) & \ldots & S(\boldsymbol{p}_{1},\boldsymbol{p}_{N_{\text{L}}}^{\prime})\\
\vdots & \ddots & \vdots\\
S(\boldsymbol{p}_{N_{\text{L}}},\boldsymbol{p}_{1}^{\prime}) & \ldots & S(\boldsymbol{p}_{N_{\text{L}}},\boldsymbol{p}_{N_{\text{L}}}^{\prime})
\end{array} & \begin{array}{ccc}
S(\boldsymbol{p}_{1},\boldsymbol{q}_{1}^{\prime}) & \ldots & S(\boldsymbol{p}_{1},\boldsymbol{q}_{N_{\text{R}}}^{\prime})\\
\vdots & \ddots & \vdots\\
S(\boldsymbol{p}_{N_{\text{L}}},\boldsymbol{q}_{1}^{\prime}) & \ldots & S(\boldsymbol{p}_{N_{\text{L}}},\boldsymbol{q}_{N_{\text{R}}}^{\prime})
\end{array}\\
\hline \begin{array}{ccc}
S(\boldsymbol{q}_{1},\boldsymbol{p}_{1}^{\prime}) & \ldots & S(\boldsymbol{q}_{1},\boldsymbol{p}_{N_{\text{L}}}^{\prime})\\
\vdots & \ddots & \vdots\\
S(\boldsymbol{q}_{N_{\text{R}}},\boldsymbol{p}_{1}^{\prime}) & \ldots & S(\boldsymbol{q}_{N_{\text{R}}},\boldsymbol{p}_{N_{\text{L}}}^{\prime})
\end{array} & \begin{array}{ccc}
S(\boldsymbol{q}_{1},\boldsymbol{q}_{1}^{\prime}) & \ldots & S(\boldsymbol{q}_{1},\boldsymbol{q}_{N_{\text{R}}}^{\prime})\\
\vdots & \ddots & \vdots\\
S(\boldsymbol{q}_{N_{\text{R}}},\boldsymbol{q}_{1}^{\prime}) & \ldots & S(\boldsymbol{q}_{N_{\text{L}}},\boldsymbol{q}_{N_{\text{R}}}^{\prime})
\end{array}
\end{array}\end{pmatrix}\ ,\label{eq:SMatrix}
\end{equation}
satisfying the unitary condition 
\[
\boldsymbol{S}(\omega)\boldsymbol{S}(\omega)^{\dagger}=\boldsymbol{S}(\omega)^{\dagger}\boldsymbol{S}(\omega)=\boldsymbol{I}_{\text{S}}
\]
where $N_{\text{S}}=N_{\text{L}}+N_{\text{R}}$ and $\boldsymbol{I}_{\text{S}}$
is the $N_{\text{S}}\times N_{\text{S}}$ scattering matrix. Equation~(\ref{eq:SMatrix})
can be expressed as: 
\[
\boldsymbol{S}(\omega)=\left(\begin{array}{ccc}
S(\boldsymbol{k}_{1},\boldsymbol{k}_{1}^{\prime}) & \ldots & S(\boldsymbol{k}_{1},\boldsymbol{k}_{N_{\text{S}}}^{\prime})\\
\vdots & \ddots & \vdots\\
S(\boldsymbol{k}_{N_{\text{S}}},\boldsymbol{k}_{1}^{\prime}) & \ldots & S(\boldsymbol{k}_{N_{\text{S}}},\boldsymbol{k}_{N_{\text{S}}}^{\prime})
\end{array}\right)
\]
where we have combined the left and right-lead wave vectors such that
for $n=1,\ldots,N_{\text{S}}$, 
\[
\boldsymbol{k}_{n}=\begin{cases}
\boldsymbol{p}_{n} & ,\ n\leq N_{\text{L}}\\
\boldsymbol{q}_{n-N_{\text{L}}} & ,\ N_{\text{L}}<n\leq N_{\text{S}}
\end{cases}\ .
\]
From Eq.~(\ref{eq:SMatrix}), we define the transition probability
matrix $\boldsymbol{W}(\omega)$ 
\begin{equation}
\boldsymbol{W}(\omega)=\left(\begin{array}{ccc}
W(\boldsymbol{k}_{1},\boldsymbol{k}_{1}^{\prime}) & \ldots & W(\boldsymbol{k}_{1},\boldsymbol{k}_{N_{\text{S}}}^{\prime})\\
\vdots & \ddots & \vdots\\
W(\boldsymbol{k}_{N_{\text{S}}},\boldsymbol{k}_{1}^{\prime}) & \ldots & W(\boldsymbol{k}_{N_{\text{S}}},\boldsymbol{k}_{N_{\text{S}}}^{\prime})
\end{array}\right)\label{eq:WMatrix}
\end{equation}
where each matrix element of $\boldsymbol{W}(\omega)$ is equal to
the absolute square of the corresponding matrix element of $\boldsymbol{S}(\omega)$,
e.g. $W(\boldsymbol{k}_{n},\boldsymbol{k}_{m}^{\prime})=|S(\boldsymbol{k}_{n},\boldsymbol{k}_{m}^{\prime})|^{2}$.

\section{Specularity parameters and transmission coefficients for zigzag-edge
graphene}

\begin{figure*}
\begin{centering}
\includegraphics[scale=0.23]{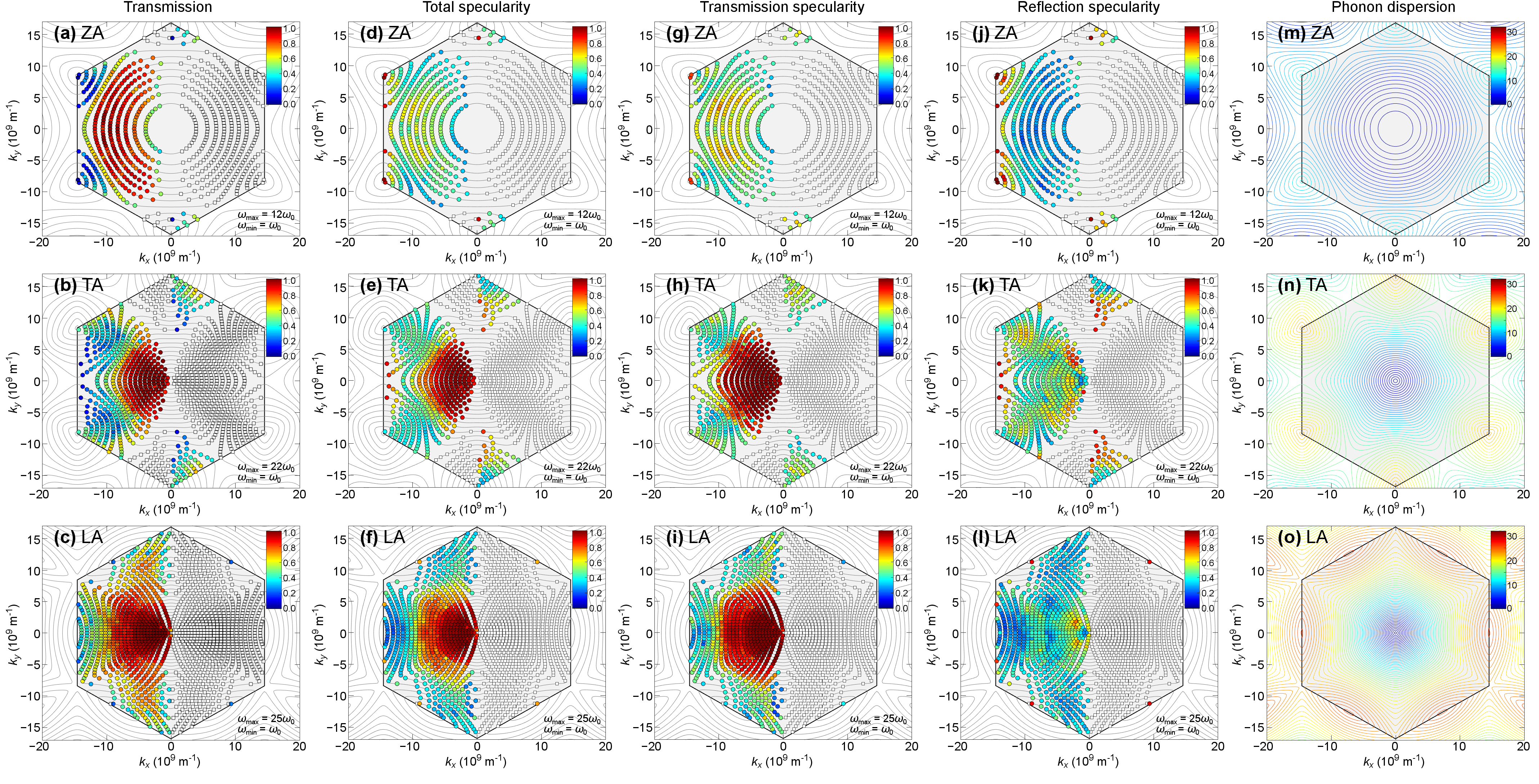}
\par\end{centering}
\caption{Mode-resolved comparison of \textbf{(a-c)} $\mathcal{T}_{n}(\omega)$,
\textbf{(d-f)} $P_{n}^{\text{total}}(\omega)$, \textbf{(g-i)} $P_{n}^{+}(\omega)$,
and \textbf{(j-l)} $P_{n}^{-}(\omega)$ for ZA, TA and LA phonons
in zigzag-edge graphene at each frequency $\omega$ point over the
frequency range $\omega=m\omega_{0}$ for $m=$ 1 to 25 and $\omega_{0}=10^{13}$
rad/s ($6.58$ meV). The isofrequency contours at each $\omega$ are
shown using solid gray lines. The modes in the incoming phonon flux
are represented by filled circles, colored according to their numerical
value as indicated in the color bars, while the modes in the outgoing
flux are represented by hollow squares. The \textbf{(m)} ZA, \textbf{(n)}
TA and \textbf{(o)} LA phonon dispersions are indicated with color
contours in intervals of $\Delta\omega=\omega_{0}/2$.}

\label{fig:ModeResolvedTransmissionAndSpecularity}
\end{figure*}

The specularity parameters (total, transmission and reflection) and
the transmission coefficients for zigzag-edge graphene in the disordered
GB model are shown in Fig.~\ref{fig:ModeResolvedTransmissionAndSpecularity}. 

\begin{figure*}
\begin{centering}
\includegraphics[scale=0.23]{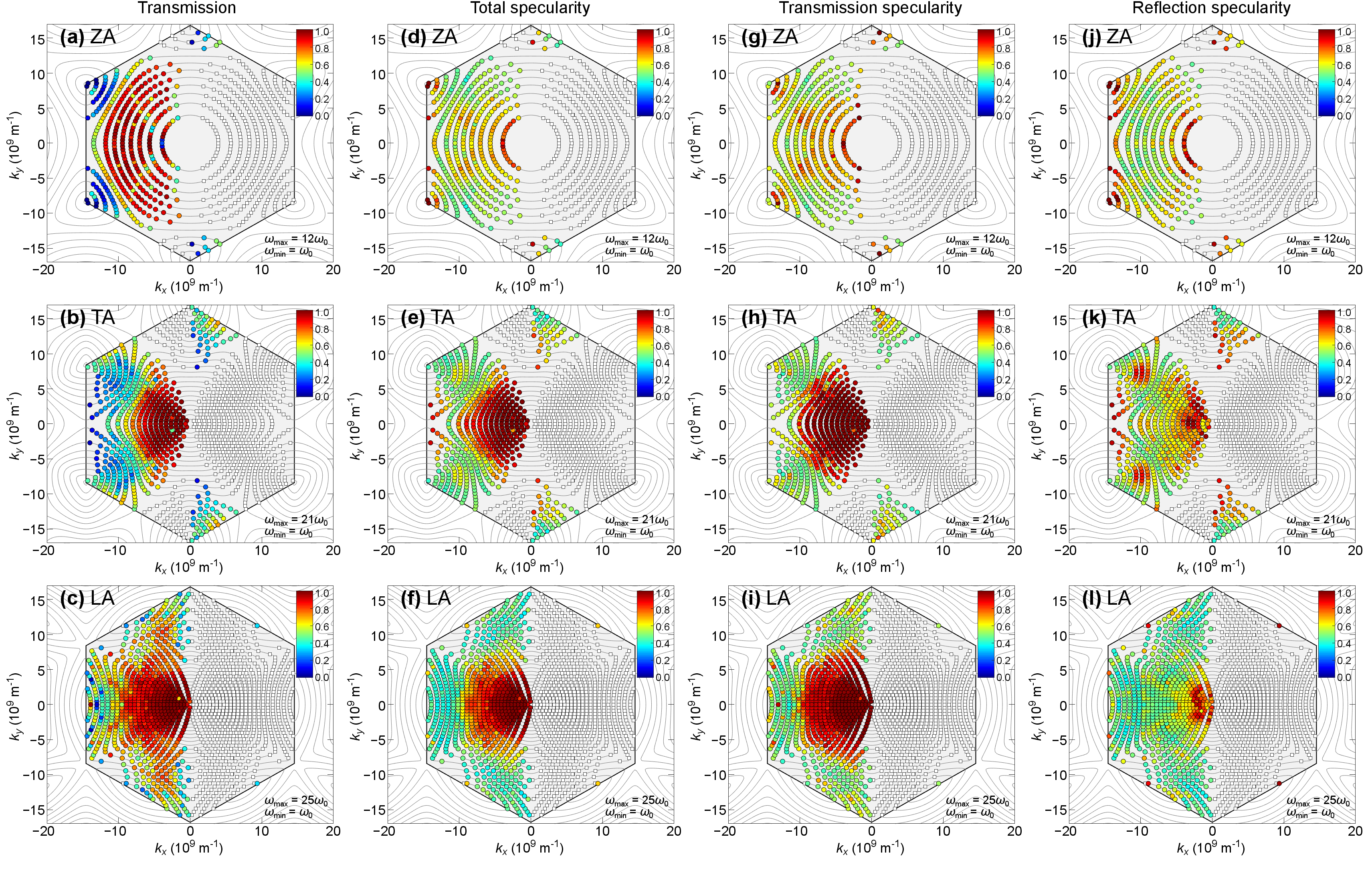}
\par\end{centering}
\caption{Mode-resolved comparison of \textbf{(a-c)} $\mathcal{T}_{n}(\omega)$,
\textbf{(d-f)} $P_{n}^{\text{total}}(\omega)$, \textbf{(g-i)} $P_{n}^{+}(\omega)$,
and \textbf{(j-l)} $P_{n}^{-}(\omega)$ for ZA, TA and LA phonons
in zigzag-edge graphene at each frequency $\omega$ point over the
frequency range $\omega=m\omega_{0}$ for $m=$ 1 to 25 and $\omega_{0}=10^{13}$
rad/s ($6.58$ meV) for a (32,32)|(56,0) graphene GB constructed from
only GB-II subunits. The isofrequency contours at each $\omega$ are
shown using solid gray lines. The modes in the incoming phonon flux
are represented by filled circles, colored according to their numerical
value as indicated in the color bars, while the modes in the outgoing
flux are represented by hollow squares. }

\label{fig:ModeResolvedTransmissionAndSpecularity_GB2}
\end{figure*}

The specularity parameters (total, transmission and reflection) and
the transmission coefficients for zigzag-edge graphene in the ordered
GB model (GB-II only) are shown in Fig.~\ref{fig:ModeResolvedTransmissionAndSpecularity_GB2}.

\section{Low-frequency ZA phonon transmission in ordered GB model (GB-II)}

\begin{figure}
\begin{centering}
\includegraphics[scale=0.28]{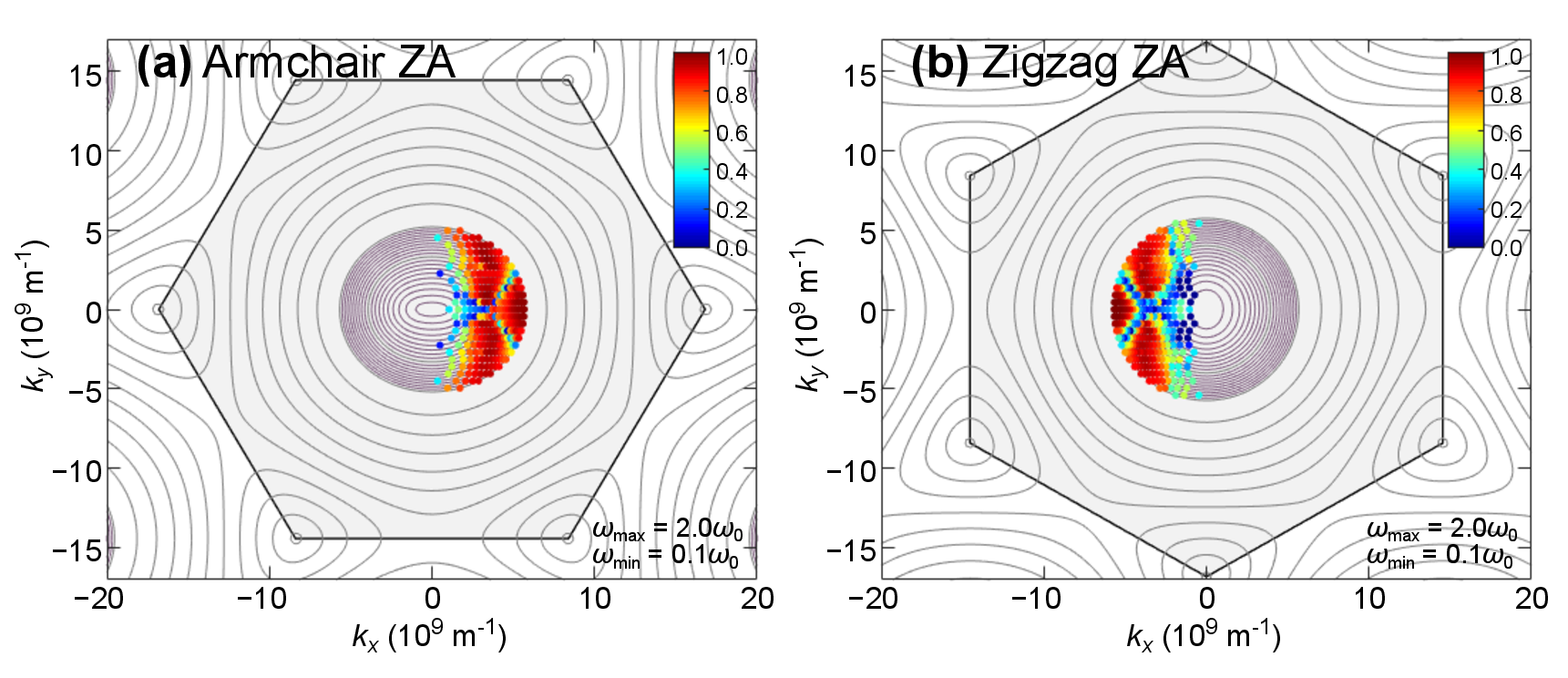}
\par\end{centering}
\caption{Mode-resolved comparison of $\mathcal{T}_{n}(\omega)$ in the low-frequency
regime for ZA phonons in \textbf{(a)} armchair and \textbf{(b)} zigzag-edge
graphene at each frequency $\omega$ point over the frequency range
$\omega=0.1\omega_{0}$ to $2\omega_{0}$ for $\omega_{0}=10^{13}$
rad/s ($6.58$ meV) with a frequency step size of $0.1\omega_{0}$
for a (32,32)|(56,0) graphene GB constructed from only GB-II subunits.
The isofrequency contours at each $\omega$ at and under $\omega=2\omega_{0}$
are displayed using solid magenta lines at intervals of $\Delta\omega=0.1\omega_{0}$
while the isofrequency contours between $\omega=2\omega_{0}$ and
$\omega=12\omega_{0}$ are displayed using solid gray lines at intervals
of $\Delta\omega=\omega_{0}$. The modes in the incoming phonon flux
are represented by filled circles, colored according to their numerical
value as indicated in the color bars.}

\label{fig:LowFreqZAPhononOrdered_GB2}
\end{figure}

In Fig.~\ref{fig:LowFreqZAPhononOrdered_GB2}, we plot the low-frequency
ZA phonon transmission coefficients in armchair and zigzag-edge graphene
for a (32,32)|(56,0) graphene GB constructed from only GB-II subunits.

\bibliographystyle{apsrev4-1}
\bibliography{PaperReferences}